\documentclass[aps,prl,reprint,superscriptaddress]{revtex4-1}

\usepackage{hyperref}
\usepackage{graphicx}
\usepackage{enumerate}
\hypersetup{
    pdfnewwindow=true,      
    colorlinks=true,       
    linkcolor=blue,          
    citecolor=blue,        
    filecolor=blue,      
    urlcolor=blue           
}

\usepackage{epsfig}
\usepackage{epstopdf}
\usepackage{subfigure}
\usepackage{amsmath}
\usepackage{amssymb}
\usepackage{amsfonts}

\usepackage{color}

\providecommand{\apj}[0]{ApJ}
\providecommand{\apjl}[0]{ApJ Lett.}

\providecommand{\aap}[0]{A\&A}
\providecommand{\aaps}[0]{Astron. Astrophys. Suppl.}

\providecommand{\mnras}[0]{MNRAS}

\providecommand{\nat}[0]{Nature}

\providecommand{\prd}{PRD}

\def\beq{\begin{equation}}
\def\eeq{\end{equation}}

\def\Mmax{M_{\max}}
\def\Psp{P_{\rm spin}}

\def\jisco{j_{\rm isco}}
\def\jeq{j_e}

\newbox\grsign \setbox\grsign=\hbox{$>$} \newdimen\grdimen \grdimen=\ht\grsign
\newbox\simlessbox \newbox\simgreatbox \newbox\simpropbox
\setbox\simgreatbox=\hbox{\raise.5ex\hbox{$>$}\llap
     {\lower.5ex\hbox{$\sim$}}}\ht1=\grdimen\dp1=0pt
\setbox\simlessbox=\hbox{\raise.5ex\hbox{$<$}\llap
     {\lower.5ex\hbox{$\sim$}}}\ht2=\grdimen\dp2=0pt
\setbox\simpropbox=\hbox{\raise.5ex\hbox{$\propto$}\llap
     {\lower.5ex\hbox{$\sim$}}}\ht2=\grdimen\dp2=0pt

\begin{document}

\title{Does the Collapse of a Supramassive Neutron Star Leave a Debris Disk?}

\author{Ben Margalit}
\email{btm2134@columbia.edu}
\affiliation{Department of Physics and Columbia Astrophysics Laboratory, Columbia University, New York, NY 10027, USA}
\author{Brian D. Metzger}
\affiliation{Department of Physics and Columbia Astrophysics Laboratory, Columbia University, New York, NY 10027, USA}
\author{Andrei M. Beloborodov}
\affiliation{Department of Physics and Columbia Astrophysics Laboratory, Columbia University, New York, NY 10027, USA}

\begin{abstract}
One possible channel for black hole formation is the collapse of a rigidly rotating massive 
neutron star as it loses its angular momentum or gains excessive mass through 
accretion. It was proposed that part of the neutron star may form a debris disk 
around the black hole. Such short-lived massive disks could be the sources of powerful 
jets emitting cosmological gamma-ray bursts. Whether the collapse creates a disk
depends on the equation of state of the neutron star. 
We survey a wide range of 
equations of states allowed by observations
and find that disk formation is unfeasible. We conclude that this channel 
of black hole formation is incapable of producing powerful jets, and
discuss implications for models of gamma-ray bursts.
\end{abstract}

\keywords{Keywords: neutron star, accretion disk, gamma-ray burst}

\maketitle


A canonical mass of neutron stars born in supernova explosions 
is $M\approx 1.4M_\odot$. The distribution of $M$ around $1.4M_\odot$ 
might, however, extend above $2M_\odot$, especially if the neutron
star is born spinning fast, with a period approaching the minimum
(``breakup'') $\Psp\sim 1$~ms.  
The additional centrifugal support allows a stable hydrostatic configuration 
with mass $M$ that would be forbidden for non-rotating stars.

Neutron stars in binary systems have additional chances to gain mass 
through accretion.
The second most massive known pulsar J1614-2230 is in a binary system and has
$M\approx 2M_\odot$  \citep{Demorest2010}. 
Its spin period is $3.15$~ms.
It is unclear if accretion is capable of spinning up the star to $\Psp\sim 1$~ms.
If this happens, the star mass may keep growing and 
remain stable even when it exceeds $2M_\odot$.

Such centrifugally supported ``supramassive'' neutron stars (SMNS) 
may also be created in mergers of neutron star binaries.
Recent observations of J1614-2230 and 
 J0348+0432 \citep{Demorest2010,Antoniadis2013}
indicate that the equation of state (EOS) of dense nuclear matter is relatively stiff, and 
therefore some mergers may initially result in a stable
object supported by pressure and fast rotation
\citep[e.g., ][]{Ozel+2010}. 
Numerical simulations show that the object
will initially rotate differentially \citep{ShibataUryu2000,RosswogDavies2002,Shibata2005,Oechslin2007,Hotokezaka2011}, but that solid body rotation will be rapidly established
following outwards transport of angular momentum via magnetic stresses and gravitational waves.  The timescale for differential rotation to be removed 
could be as short as tens of ms \citep{ShibataTaniguchi2006}, and
will almost certainly be much shorter than 10~s (e.g., \citep{Shapiro2000}).  
The heat stored in the merger product is also mostly lost to neutrino emission within seconds (e.g., \citep{BurrowsLattimer86}).

The SMNS is fated to collapse to a black hole. 
Its lifetime is controlled by the eventual loss of
angular momentum (spindown-induced collapse) or excessive mass growth 
(accretion-induced collapse). 
The collapse is associated with a huge release of gravitational energy
and could produce a bright 
transient event --- a burst of electromagnetic radiation, such as 
a cosmological gamma-ray burst (GRB).

This GRB trigger is plausible 
if the equatorial part of the neutron star is not 
immediately swallowed by the black hole but forms a compact, massive,
centrifugally supported disk around it.
Jets of hot plasma and radiation are expected to emerge from 
the debris disk and power the burst (e.g., \citep{Narayan+1992}).

In the merger scenario, the SMNS eventually collapses due to its gradual spindown, 
which removes the rotational support in minutes to hours.
The spindown timescale depends on the magnetic field of the merger product, 
which is likely amplified to $B\sim 10^{15}$~G during the merger \citep{ThompsonDuncan1993,Giacomazzo+14}.
This implies a moderate delay of the collapse-powered burst following the gravitational 
waves that are emitted during the merger and hopefully detected by Advanced LIGO \citep{RezzollaKumar2015,CiolfiSiegel2015}.

The goal of this Letter is to assess if the key condition for this burst scenario --- 
a massive debris disk after the collapse --- can be satisfied. The structure
of the SMNS and hence the outcome of its collapse are controlled by the EOS of the dense nuclear matter $P(\rho)$.
Available
general relativistic simulations of the collapse 
do not show disk formation \citep{Shibata2003,Baiotti2005}.
These simulations, however, implemented only 
simplified EOS.
In particular,
\cite{Shibata2003} 
used the
polytropic 
$P\propto \rho^{1+1/n}$ with index
$n \leq 2$,
and found
that less than $10^{-3} M_{\odot}$ 
remains outside the black hole at the termination of the simulation, comparable to their numerical resolution. They also 
found
that for an extremely soft EOS (with $n=2.9$ 
and
$3$) 
disks
can form, 
however such EOS
are incompatible with 
observations of neutron stars. The remaining open possibility is that 
a different form of the EOS could lead to disk formation, e.g. soft at high densities
(which gives a compact inner core --- the seed for a future black hole) and 
stiff at lower densities (which gives an extended outer core with a high angular 
momentum). 

In this Letter we explore a wide range of EOS in search for one
that could possibly give a debris disk.  Instead of carrying out full-fledged and computationally expensive hydrodynamic simulations of SMNS collapse, we employ a simple method. 
We analyze
the equilibrium 
hydrostatic 
configuration prior to 
the
collapse
and check if it satisfies a necessary condition 
for formation of a debris disk after the collapse.


{\it Condition for Disk Formation.}$-$
A stringent criterion on disk formation can be derived by assuming that all but an infinitesimal amount of the SMNS's mass and angular momentum are inherited by the newly formed Kerr black hole. Matter at the SMNS equator has the largest specific angular momentum, $j_e$, and hence is the most likely to comprise a disk.    
The angular momentum is conserved during collapse, as long as
magnetic and viscous torques are negligible and the spacetime remains 
axisymmetric. The centrifugal barrier will stop the equatorial matter from plunging the horizon if 
$\jeq$ exceeds the specific angular momentum of the inner-most stable circular orbit (ISCO)
in the Kerr metric of the nascent black hole,
\begin{equation} \label{eq:j_criteria}
  \jeq >\jisco(a) ~~~ \Rightarrow ~~~ \text{disk formation is possible} .
\end{equation}
Note that $\jisco$ depends on the spin parameter $a=Jc/GM^2$ where 
$J$ is  the angular momentum inherited by the black hole from the SMNS.
A similar criterion has been employed previously to the collapse of supermassive gas clouds \citep{ShapiroShibata2002}.

\medskip


{\it Maximally Rotating Maximal Mass.}$-$
We construct axisymmetric neutron star models using the \texttt{rns} code \citep{StergioulasFriedman1995, NozawaStergioulas1998}, which calculates relativistic rotating 
hydrostatic equilibria following the method outlined in \cite{Cook1994a,Cook1994b}. 
The collapse occurs when the stellar mass exceeds $\Mmax$ at which the star
becomes secularly unstable \citep{Friedman1988} and no hydrostatic solution is found.   

$\Mmax$ depends on the angular momentum $J$ and the EOS of dense 
nuclear matter. For a given EOS, we calculate $\Mmax(J)$ and find $a$ and $\jeq$ 
immediately prior to 
collapse.  Disk formation is clearly impossible for a non-rotating star because matter will fall radially into the newly-formed Schwarzschild black hole.  
As 
$J$ and hence $j_e$ are increased, black hole spin $a$ 
increases
and hence $j_{\text{isco}}$ decreases.  
Condition~(\ref{eq:j_criteria})
could thus in principle be 
satisfied
at some point along the maximal mass sequence.

The maximal mass sequence 
$\Mmax(J)$ cannot be extended indefinitely as it eventually
reaches the mass-shedding limit, beyond which the 
co-rotating orbital frequency at the SMNS equator exceeds the 
SMNS rotation frequency. 
This point defines the maximally rotating maximum mass (MRMM), 
$\Mmax(J_{\max})$, which is typically 10-30\% higher than $\Mmax(0)$. 
The collapsing MRMM has the best chance to form a debris disk
but this is not guaranteed. Although $\jeq$ of the MRMM 
is just sufficient to orbit the hydrostatic star, the spacetime metric
changes after the collapse and the same $\jeq$ can fail to sustain Keplerian rotation around 
the nascent black hole.
If condition~(\ref{eq:j_criteria}) is
not met for the MRMM, it will not be met for any
slower rotating maximal mass models and we may conclude that disk formation is 
impossible for this EOS.

The input parameters of the rns code are the central energy density and 
the oblateness of the star. For a given oblateness we find the maximal mass model
by varying the central energy density. 
Then we step along the maximum mass sequence toward MRMM by increasing the 
oblateness parameter. At the end of the sequence we iterate the oblatness
until the mass shed limit is found to within a specified accuracy.  
At each step we check if the disk formation criterion~(\ref{eq:j_criteria}) is satisfied.

\begin{figure}
\epsfig{file=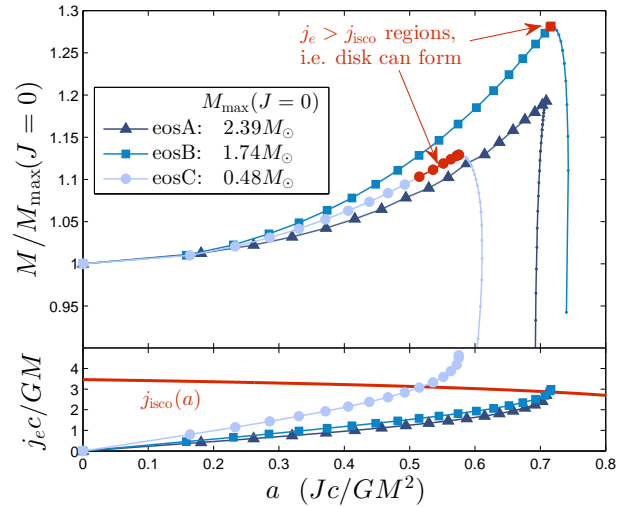,angle=0,width=0.5\textwidth}
\caption{Sequence of neutron star mass $M$ and spin parameter $a$ for three sample EOSs, illustrating our method for assessing the possibility of disk formation following SMNS collapse. The top portion of the figure shows the maximal mass sequence (triangles/squares/circles) and mass shed limits (small points) for each EOS.  Masses are normalized to the maximal value for a non-rotating star corresponding to each EOS.  The bottom portion shows the (dimensionless) 
specific angular momentum of a test particle at the SMNS equator, 
$\jeq c/GM$,
along the maximal sequence curves. 
A solid red line denotes the minimal angular momentum required to orbit the resulting Kerr black hole with spin parameter $a$, $j_\mathrm{isco}(a)$. According to the criterion~(\ref{eq:j_criteria}), disk formation is ruled out as long as $j_e$ lies below this red curve.   
These three EOS are also marked in Figure~\ref{fig:EOS_Phase_Diagram}.} \label{fig:MJ_Diagram}
\end{figure}

Figure \ref{fig:MJ_Diagram} illustrates our procedure for three representative EOS, labeled eosA, eosB, and eosC, respectively.  
For eosA, $j_e < j_{\rm isco}$ for all black hole spin $a$, so disk formation is impossible according to criterion~(\ref{eq:j_criteria}).  For eosC, $j_e > j_{\rm isco}$ for $a \gtrsim 0.5$, indicating that a disk could form; however, the maximum non-rotating mass for this unrealistically soft EOS is only $0.48 M_{\odot}$.  Disk formation is also possible for eosB, but only for a very 
narrow range of $J$
near the mass-shedding limit.  

\medskip

{\it Survey of the EOS Space.}$-$
The possibility of disk formation
is controlled by the high density EOS, which is poorly known.
Therefore, below
we conduct a 
survey over a broad range of EOS.
Our goal is
to check whether 
it is possible to
simultaneously satisfy the disk 
formation
criterion and current observational constraints on neutron star radii and masses.

We parametrize the EOS 
at $\rho>\rho_0 = 10^{14.3}$~g~cm$^{-3}$
as a broken power law. This choice is motivated by previous works \citep{Read2009} which 
show that a piecewise polytrope can reliably reproduce a variety of EOS models.
The break is fixed at density
 $\rho_1 = 10^{14.7} ~\text{g} ~\text{cm}^{-3}$.  
At densities below $\rho_0$
we use the SLy EOS \citep{DouchinHaensel2001} 
with the approximation
of \cite{Read2009}, and we fix $P(\rho_0)$ to the SLy value.  

With fixed $\rho_1$
we are left with only two free parameters:
$P_1=P(\rho_1)$
and the 
power-law index 
 at $\rho>\rho_1$, $\Gamma_2=d\ln P/d\ln\rho$. Two degrees of freedom
in the EOS may be insufficient to predict
observables to within $\sim 1\%$ accuracy (e.g., as in \citep{Read2009}).
However, this form of EOS is sufficiently flexible for our purposes,
allowing independent variation of the SMNS mass 
$M$ 
and radius
$R$. These parameters
determine the 
star's
compactness
$M/R$, the key
factor for disk formation.

\begin{figure}
\epsfig{file=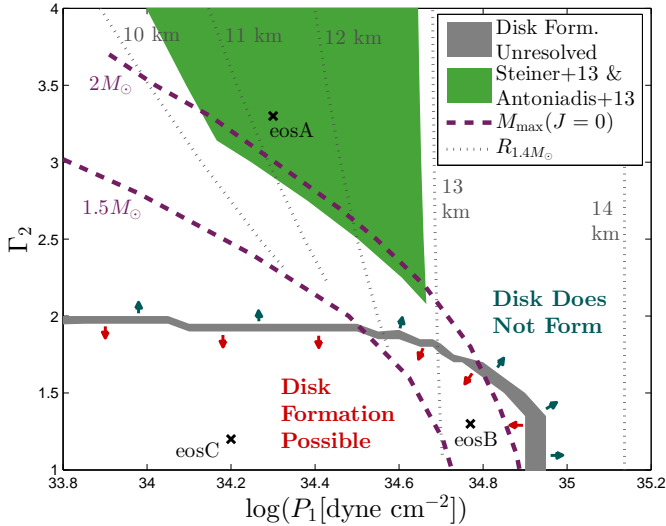,angle=0,width=0.5\textwidth}
\caption{Regions of allowed and forbidden disk formation in the EOS parameter space. 
Dashed purple 
curves
show 
contours
of constant maximum mass for non-rotating neutron stars, 
while dotted black lines indicate constant radius values for a $1.4 M_{\odot}$ non-rotating star. The green region shows the 2$\sigma$ allowed parameter space based on observed neutron star masses \citep{Antoniadis2013} (bottom boundary) and constraints on neutron star radius \citep{Steiner2013} (left and right side boundaries).
} \label{fig:EOS_Phase_Diagram}
\end{figure}

The results of our numerical survey of the parameter space $P_1$-$\Gamma_2$ are 
shown in Figure~\ref{fig:EOS_Phase_Diagram}. For ``stiff'' EOS above the grey strip even 
the MRMM configuration fails to meet the criterion~(\ref{eq:j_criteria}), and thus disk 
formation is ruled out. The criterion is met by the MRMM below the grey strip (and possibly 
inside the strip where it is numerically unresolved).

Small $P_1$ or $\Gamma_2$ values are however problematic as they predict low $\Mmax$ while observations 
demonstrate the existence of neutron stars with $M\approx 2M_\odot$ \citep{Demorest2010,Antoniadis2013}, 
even at moderate rotation when centrifugal effects may be neglected (the $39$~ms spin period of J0348+043 is slow enough that it can be treated as essentially non-rotating for the purpose of constraining the maximal neutron star mass).

An additional observationaly accessible parameter is the radius of normal neutron stars with moderate rotation and canonical mass $M\approx 1.4M_\odot$. For instance using observations of transiently accreting and bursting neutron stars \citep{Steiner2013} reported $R_{1.4 M_{\odot}} = 10.42-12.89$~km at 2$\sigma$. We note that current neutron star radius constraints are subject to uncertainties in both astrophysics and nuclear physics modeling and the radius constraints are not entirely settled yet (cf. e.g., \cite{Suleimanov+11,Guillot+13}).

For any candidate 
EOS one should check its prediction for $\Mmax(J\approx 0)$ as well as $R_{1.4 M_{\odot}}$, which can be tested against observations.
Figure~\ref{fig:EOS_Phase_Diagram} 
shows the contours
of constant 
$M_\mathrm{max}(J=0)$ and 
$R_{1.4 M_{\odot}}$
on the $P_1$-$\Gamma_2$ plane together with the observational constraints.
The condition $M_{\max}>2M_\odot$ alone excludes almost the entire region where disk formation is possible. A significant gap appears between this region and the allowed region if following \cite{Steiner2013} we also require $R_{1.4M_\odot}<13$~km.
We also note that even below the grey strip 
formation of a debris disk requires significant fine-tuning toward the MRMM configuration.
Disk formation quickly becomes impossible if $J$ is reduced below $J_{\max}$ 
(see Figure~\ref{fig:MJ_Diagram}, in particular model eosB).

\medskip


{\it Discussion and Astrophysical Implications.}$-$
Our method employs a simple parametrization for the high density EOS as a piecewise polytrope, and hence may not replicate nuances of realistic EOSs (for instance, \cite{Read2009} shows that it tends to overestimate the sound speed).  This parametrization is, however, sufficient to 
capture 
the overall mass distribution of the star, which is most important to our analysis.
Since our results in Figure~\ref{fig:EOS_Phase_Diagram} show a significant gap between the observationally allowed and the disk forming regions, any analysis using a more complex parametrization (as in e.g. \cite{Read2009,Steiner2013}) is likely to yield a similar conclusion. This is especially so when considering that much of the formally allowed disk formation region will not produce a disk in practice without fine tuning of the SMNS angular momentum.  

In particular, we have applied our method directly to the entire list of parametrized EOSs given in \cite{Read2009} (their Table III), and find that none of these support disk formation. Interestingly, the robustness of our main conclusion 
relies
in part on the recent discovery of a $2M_{\odot}$ neutron star \citep{Demorest2010,Antoniadis2013} and hence could not have been made with as much confidence prior to 2010, when the largest known mass was 1.74$\pm 0.04 M_{\odot}$.

Although lower limits on the maximum neutron star mass are well
established by dynamical measurements, 
observational constraints on the neutron star radius are subject to
systematic uncertainties 
\citep[e.g.,][]{Miller13}.  It is thus important to note that our
conclusion that disk formation is unlikely depends most sensitively on the
established maximum mass constraints, and less critically on the neutron star
radius.

Our analysis assumed axisymmetric collapse.
This is reasonable since non-axisymmetric perturbations will likely be damped out via gravitational waves. Furthermore, if the amount of surviving disk mass is 
determined by deviations from axisymmetry
then producing a disk of 
an interesting
mass $\gtrsim 10^{-3} M_{\odot}$ 
translates into a radial perturbation of $\gtrsim 2 ~\mathrm{km}$, an unlikely occurrence.

We have additionally assumed that 
magnetic or viscous torques
do not affect the SMNS matter during the collapse.  
Numerical hydrodynamical simulations 
consistently 
show
that the SMNS matter collapses on a dynamical timescale with 
approximate conservation of angular momentum and
negligible 
dissipation 
effects on fluid streamlines
\citep{Shibata2003}.
Magnetic fields could become dynamically important only when they are extremely strong.
Such fields could also slightly affect the SMNS structure. Its radius would be increased
up to $\sim 16 \%$ in the most extreme case of magnetic pressure equal to thermal pressure (e.g., \citep{Kamiab2015}). 

Our results have implications for some GRB models.
Electromagnetic emission from 
SMNS formed in
neutron star binary mergers has been proposed by many authors \citep[e.g.][]{Metzger2008,Bucciantini2012,Rowlinson2013,Gompertz+15} to explain long-lived X-ray flares (``extended emission") and plateaus observed following short duration GRBs \citep[e.g.][]{NorrisBonnell2006,Nousek2006,Zhang2006}, which in some cases have been observed to terminate abruptly in a way suggesting a SMNS that has collapsed to a black hole \citep{Rowlinson+10}. 
These magnetar models have been criticized because it is not clear how to produce the relativistic jet responsible for the initial GRB itself as the result of baryonic pollution from the young neutron star remnant \citep[e.g.,][]{Murguia-Berthier+14}. This has recently led to the suggestion of a 
``Time Reversal" scenario \citep{RezzollaKumar2015,CiolfiSiegel2015}, whereby black hole formation and the GRB is delayed for tens or hundreds of seconds following the merger, but due to light time travel effects is observed before X-rays from the SMNS remnant cease.  A similar physical situation, which posits the collapse of a SMNS to a black hole following the accretion of matter from a binary companion (accretion-induced collapse; e.g., \citep{Macfadyen2005,GiacomazzoPerna2012}) is also commonly invoked as an alternative to neutron star merger models for short GRBs.

Both these alluring models (accretion-induced collapse and Time Reversal) require a debris disk after the SMNS collapse in order to power the short GRB. Our results show that this assumption contradicts the stiff nuclear EOS inferred from observations of neutron stars.

This does not necessarily mean that 
SMNS
collapse will have no observational electromagnetic signature. For instance \cite{Lehner2012,Falcke&Rezzolla13} suggest that if the SMNS is initially magnetized, a significant electromagnetic transient could arise regardless of any surrounding accretion disk.  However, such a transient is unlikely to last 
many dynamical times across the black hole horizon and hence 
may fail to
explain the 
$0.1-1$~s
duration of observed short GRBs.  

Our model assumes solid body rotation and a cold EOS and hence does not
rule out a disk if the black hole forms shortly following a binary neutron star
merger.  Disk formation in fact appears to be a robust outcome of general
relativistic simulations of the merger process \citep[e.g.,][]{ShibataTaniguchi2006}.  Thermal
pressure is only sustained for a few seconds after the merger, until
neutrino cooling sets in.  More importantly, the merger remnant is
primarily supported by {\it differential} rotation, such that the collapse
is usually initiated by the outwards redistribution of angular momentum,
as is expected to occur on a timescale of tens or hundreds of milliseconds
due to magnetic or viscous stresses. 
Since in this case collapse occurs prior to the establishment of solid body rotation throughout the remnant, disk formation is much more likely than in the case of a delayed collapse.

Finally, our results also render untenable proposed scenarios 
for long duration GRBs which postulate 
a long delay (exceeding hours or days) between the core collapse of a massive star and the formation of a black hole 
with a debris accretion disk
\citep{VietriStella1998}.

{\it Acknowledgments.}$-$We acknowledge helpful conversations with Luis Lehner, Pawan Kumar, Luciano Rezzolla, and Andrew Steiner.  B.D.M. gratefully acknowledges support from NASA {\it Fermi} grant NNX14AQ68G, NSF grant AST-1410950, and the Alfred P. Sloan Foundation. GRB research of A.M.B. is supported by NSF grant AST-1412485, NASA grants  NNX14AI94G and NNX15AE26G.

\bibliography{Disk_Formation_Bibliography}

\end{document}